\documentclass[twocolumn,pre,showpacs,superscriptaddress]{revtex4}
\usepackage{graphicx}%
\usepackage{color} 
\usepackage{transparent}
\usepackage{psfrag}
\usepackage{dcolumn}
\usepackage{amsmath}
\usepackage{amssymb}
\usepackage{latexsym}
\usepackage{hyperref} 
\usepackage{booktabs}
\usepackage{latexsym}
\begin{document}

\def\bea{\begin{eqnarray}}
\def\eea{\end{eqnarray}}
\def\beq{\begin{equation}}
\def\eeq{\end{equation}}
\def\f{\frac}
\def\k{\kappa}
\def\e{\epsilon}
\def\ve{\varepsilon}
\def\be{\beta}
\def\D{\Delta}
\def\h{\theta}
\def\t{\tau}
\def\a{\alpha}

\def\cDa{{\cal D}[X]}
\def\cD{{\cal D}[x]}
\def\cL{{\cal L}}
\def\cLo{{\cal L}_0}
\def\cLa{{\cal L}_1}

\def\Re{{\rm Re}}
\def\sj{\sum_{j=1}^2}
\def\rk{\rho^{ (k) }}
\def\rek{\rho^{ (1) }}
\def\cek{C^{ (1) }}
\def\rz{\rho^{ (0) }}
\def\rt{\rho^{ (2) }}
\def\rtb{\bar \rho^{ (2) }}
\def\trk{\tilde\rho^{ (k) }}
\def\trek{\tilde\rho^{ (1) }}
\def\trz{\tilde\rho^{ (0) }}
\def\trt{\tilde\rho^{ (2) }}
\def\r{\rho}
\def\tD{\tilde {D}}

\def\s{\sigma}
\def\kb{k_B}
\def\bF{\bar{\cal F}}
\def\F{{\cal F}}
\def\la{\langle}
\def\ra{\rangle}
\def\nn{\nonumber}
\def\up{\uparrow}
\def\dn{\downarrow}
\def\S{\Sigma}
\def\dg{\dagger}
\def\d{\delta}
\def\p{\partial}
\def\l{\lambda}
\def\L{\Lambda}
\def\G{\Gamma}
\def\o{\Omega}
\def\w{\omega}
\def\g{\gamma}

\def\jv{ {\bf j}}
\def\jr{ {\bf j}_r}
\def\jd{ {\bf j}_d}
\def\jdd{ { j}_d}
\def\noi{\noindent}
\def\a{\alpha}
\def\d{\delta}
\def\p{\partial} 

\def\la{\langle}
\def\ra{\rangle}
\def\e{\epsilon}
\def\n{\eta}
\def\g{\gamma}
\def\break#1{\pagebreak \vspace*{#1}}
\def\hf{\frac{1}{2}}

\title{Entropy production by active particles: Coupling of odd and even functions of velocity}
\author{Debasish Chaudhuri}
\email{debc@iopb.res.in}
\affiliation{Institute of Physics, Sachivalaya Marg, Bhubaneswar 751005, India
}
\affiliation{Homi Bhabha National Institute,   Anushaktinagar, Mumbai  400094, India}

\date{\today}

\begin{abstract}
Non-equilibrium stochastic dynamics of several active Brownian systems are modeled in terms of non-linear velocity dependent force. In general, this force may consist of both even and odd functions of velocity.  We derive the expression for total entropy production in such systems using the Fokker-Planck equation. The result is consistent with the expression for stochastic entropy production in the reservoir, that we obtain from  probabilities of time-forward and time-reversed trajectories, leading to fluctuation theorems. Numerical simulation is used to find probability distribution of entropy production, which shows good agreement with the detailed fluctuation theorem.  %
\end{abstract}
\pacs{05.40.-a, 05.40.Jc, 05.70.-a}

\maketitle

\section{Introduction}
Active particles are self propelled entities that perform locomotion utilizing internal energy, even in the absence of an external driving force. 
The internal energy source may be replenished by food, e.g., in animal to bacteria, or local chemical fuel in the form of ATP in molecular 
motors. Studies of active particles have been motivated by dynamic cluster formation in birds, fish, or animal~\cite{Vicsek2012, Vicsek1995},
active Brownian motion of self propelled colloids or nano rotors~\cite{Howse2007,Zheng2013,Nourhani2013}, and even by the motion of vibrated 
granular systems~\cite{Feitosa2004, Joubaud2012,Kumar2011}. The self propelled motion of several active Brownian particle (ABP) systems
may be described in terms of a non-linear velocity dependent force~\cite{Romanczuk2012, Badoual2002, Julicher1995}. 

A simple example of non-linear velocity dependent force is the motion of a projectile through a compressible fluid. A particle of velocity $v$ displaces a volume of fluid proportional to $v$, thus imparting a 
change in momentum proportional to $v^2$ in the medium per unit time. The particle in turn encounters an equal and opposite force, which is an even function of $v$ but directed opposite to the direction of motion.    In an active system, on the other hand, non-linear velocity dependent force may support the motion at small velocities. Two such models are the  Rayleigh-Helmholtz model~\cite{Romanczuk2012}, and the energy depot model~\cite{Schweitzer1998, Zhang2008}.

Systems with small degrees of freedom (dof), and driven arbitrarily out of equilibrium are describable within the framework of stochastic 
thermodynamics~\cite{Sekimoto1998,Bustamante2005,Seifert2012}. This uses stochastic counterparts of  thermodynamic observables like work, entropy etc.
The detailed fluctuation theorem imposes strict symmetry to the probability distribution of entropy production in passive Brownian systems driven out of equilibrium, e.g., 
small assembly of nano-particles, colloids, granular matter, and
polymers~\cite{Jarzynski2011,Jarzynski1997,Crooks1999,Wang2002,Liphardt2002,Feitosa2004,Narayan2004,Kurchan2007}. 
Although stochastic entropy production (EP) can be negative, probability of such events is exponentially suppressed with respect to positive entropy producing trajectories~\cite{Evans1993,Gallavotti1995,Lebowitz1999, Seifert2005,Seifert2008}. 
Stochastic thermodynamics of dry friction has been considered recently~\cite{Gnoli2013, Sarracino2013,Cerino2015}. In the context of coarse grained theories, it is known that simplification of a model by integrating out faster dofs leads to loss of information and EP~\cite{Puglisi2010,Crisanti2012,Mehl2012a, Jia2016}.

Several experiments on  colloids and granular matter were used to verify fluctuation theorems~\cite{Wang2002, Blickle2006, Speck2007, Joubaud2012}. Using Jarzynski equality, the  free energy landscape of 
 RNA was obtained from distribution of non-equilibrium work done~\cite{Liphardt2002,Collin2005}. 
Fluctuation theorems have been derived for models of molecular motors as well~\cite{Seifert2011,Lacoste2011,Lacoste2009}. Autonomous torque generation by
rotary motor was estimated applying detailed fluctuation theorem on stochastic trajectories~\cite{Hayashi2010,Hayashi2012}. 
Stochastic thermodynamic description of   the  Rayleigh-Helmholtz  and energy depot model were obtained recently~\cite{Ganguly2013, Chaudhuri2014}. 

In this paper, we study stochastic thermodynamics for ABPs in the presence of general velocity dependent forces containing both odd and even functions of velocity. 
Unlike the Rayleigh-Helmholtz model,
the presence of an even function of velocity, and its coupling with the odd function leads to EP in velocity space even in the absence of external force or potential.
Using the Fokker-Planck equation, we derive the expression for total EP in the reservoir. The result is consistent with the expression for stochastic EP
that we find independently from the probability distributions of time forward and time reversed trajectories. This gives us several {\em excess} entropy terms, in addition to
Clausius like dependence of stochastic EP on stochastic heat flux. 
We further discuss the amount of loss of EP inherent to a coarse grained model of ABP, like the Rayleigh-Helmholtz model, 
with self propulsion in absence of a mechanism behind it, by explicitly considering an energy depot like mechanism producing activity.
The path probability calculations of the ABP model lead to detailed and integral fluctuation theorems (FT) for EP. Finally, we use numerical 
simulations to find the probability distribution of EP that shows good agreement with the detailed FT.

\section{Non-linear velocity dependent force}

The dynamics of this ABP under non-linear velocity dependent forces $F(v) = \zeta(v) + \xi(v)$ such that $\zeta(-v) = -\zeta(v)$ and $\xi(-v)=\xi(v)$ 
is described by the Langevin equations of motion
\bea
\dot x &=& v \nn\\
\dot v &=&   \eta(t) + g(v) + \xi (v) - \p_x U(x) + f(t).
\label{lange}
\eea 
where $g(v) = - \g v + \zeta(v)$ denotes an odd function of velocity with $-\g v $ the viscous dissipation due to surrounding environment,
$\eta(t)$ is the Gaussian white noise obeying $\la \eta(t) \ra =0$, $\la \eta(t) \eta (t')\ra = 2  D_0 \d(t-t')$ with $D_0=\g \kb T$, 
with $\kb$ the Boltzmann constant, and $T$ is the temperature of surrounding heat bath. 
$U(x)$ is an external potential, and $f(t)$ a time-dependent control force. 
We use particle mass $m=1$ throughout this paper. 

The  Fokker-Planck equation corresponding to Eq.(\ref{lange}) is given by 
\bea
\p_t P(x,v,t) &=& -\p_x (v P) - \p_v \left[ \left( g(v) + \xi(v)  + \bF  \right) P \right] \crcr 
&& + D_0 \p_v^2 P \equiv - \nabla. \jv
\label{fp_eq}
\eea
where $ \nabla = (\p_x, \p_v)$, $g(v)$ and $\xi(v)$ are odd and even functions of velocity, respectively, and $\bF = f(t) - \p_x U$. 
Under time reversal, position $x$ is an even variable, and velocity $v$ is an odd variable.  
The probability current  $\jv = \jr + \jd$ with $\jr = [v P, (\bF + \xi(v))\, P]$ the time-reversal symmetric part, and $\jd =  (0, g(v) P - D_0 \p_v P)$ the dissipative part.
Note that when $\jd =(0,0)$, FP equation remains invariant under time reversal. Whereas, if $\jr = (0,0)$, the right hand side of FP equation picks up an overall negative sign. 
The presence of dissipative current $\jd$ denotes breaking of time-reversal symmetry and entropy production (EP).

The model presented here should be interpreted as a coarse grained model of self propulsion, incorporating an internal energy source for each particle. Assuming a time
scale separation of the internal degrees of freedom (dof) with respect to the relatively {\em slow} mechanical motion of the particles, one can integrate out these fast internal dof. An assumption of steady state for these internal dof allows one to effectively incorporate them via a velocity dependent force in mechanical motion~\cite{Schweitzer1998}. Note that this force renders an inherently non-equilibrium nature to the ABPs. Even in a special case of detailed balance in the mechanical dof for $\xi(v)=0$, this dissipative probability current from internal energy source to mechanical motion leaves the particle out of equilibrium, a fact reflected in their non-Gaussian steady state velocity distribution~\cite{Chaudhuri2014}.

\section{Entropy production}
\label{detailed}
\subsection{From the Fokker-Planck equation}
\label{ep_fp1}
We first calculate EP using the FP equation. 
The definition of non-equilibrium Gibbs entropy  $S(t) = -\kb \int dx\, dv P(x,v,t) \ln P(x,v,t)$, along with the FP equation, may be used to obtain the rate of EP,
\bea
\f{d S}{d t} &=& -\kb \int dx\, dv\, \ln P\, \f{\p P}{\p t} \crcr
&=& \kb \int dx\, dv\, \ln P\,  [ \nabla \cdot (\jr + \jd) ]. 
\eea
In obtaining the first step, we used the normalization condition of $P$ that leads to $\int dx\, dv\, \p_t P = 0$. 
Integration by parts twice, 
\begin{align}
& \int dx\, dv\, \ln P\, \nabla \cdot \jr 
 = \int dx\, dv\, P\, \nabla \cdot (\jr/P) 
= \la \p_v \xi(v) \ra \nn
\end{align}
using $\jr / P = [v, (\bF + \xi(v))]$ in the last step. The integral involving dissipative current, leads to
\begin{align}
& \int dx\, dv\, \ln P\, \nabla \cdot \jd 
 =  - \int dx\, dv\, \jdd \, \f{g(v)-\jdd/P}{D_0}. \nn
\end{align}
In deriving the above relation,
we used the expression of the velocity component of dissipative current 
$\jdd = g(v)P - D_0 \p_v P$ to write $\p_v \ln P$ in terms of $\jdd$. Thus,
\bea
\f{1}{\kb}\f{d S}{d t} &=& \la \p_v \xi(v) \ra + \int dx\, dv \f{\jdd^2}{P\, D_0} - \f{1}{D_0}  \int dx\, dv\, \jdd g(v) \nn.
\eea
This leads to the total EP 
\bea
\dot S_t = \dot S + \dot S_r = \kb \int dx\, dv \f{\jdd^2}{P\, D_0} \geq 0
\label{stdot}
\eea
in agreement with the second law of thermodynamics. This is characterized by 
the dissipative non-equilibrium processes in the system in terms of $\jdd$. 
The entropy flux to reservoir is the same as the EP in reservoir 
\bea
\f{1}{\kb}\dot S_r = -\la \p_v \xi(v) \ra + \f{1}{D_0}  \int dx\, dv\, \jdd g(v).
\label{Srdot}
\eea

The definition $S(t) = -\kb \int dx\, dv P \ln P$ leads to the definition of stochastic entropy in the system $s(t) = -\kb \ln P(x,v,t)$ such that
$S(t) = \la s(t) \ra$~\cite{Seifert2005}. Similarly the stochastic EP in reservoir $\dot s_r$ is expected to obey  
$\dot S_r = \la \dot s_r \ra$. 
The thermodynamic average of stochastic quantities involve a two step averaging, (i)~over trajectories, 
(ii)~over phase space with probability $P(x,v,t)$~\cite{Seifert2005}.
Let us obtain an expression for stochastic EP $\dot s_r$ in reservoir by {\em undoing} these averaging from the expression of $\dot S_r$ given
in Eq.(\ref{Srdot}). Removing the averaging over phase space with probability $P$ suggests a form $- \p_v \xi(v) + [\jdd \, g(v)/P D_0]$ for $\dot s_r/\kb$.
Note that the velocity component of probability current $j_v = [\bF + \xi(v)]P + \jdd$. The velocity current is related to particle velocity by the averaging over 
stochastic trajectories, $ \la \dot v | x,v,t \ra = j_v/P = [\bF + \xi(v)] + \jdd/P$. Removing the averaging over stochastic trajectories, suggests replacing 
$\jdd/P$ by $\dot v - [\bF + \xi(v)] $. Thus the stochastic expression for EP in the reservoir can be written as
\bea
\f{1}{\kb}\dot s_r &=& - \p_v \xi(v) + \f{1}{D_0} g(v) [ \dot v - (\bF + \xi(v)) ] \nn\\
&=& - \p_v \xi(v) + \f{g(v)}{D_0}  [ \dot v  + \p_x U - f(t) - \xi(v) ]. 
\label{sr_dot}
\eea
It is not immediately clear whether performing such undoing of integrations over stochastic trajectories and probability distributions indeed 
is a natural way to obtain the stochastic  EP of the reservoir. The same thermodynamic expression may result from various other
stochastic definitions, if the excess stochastic terms cancel out after averaging.
Thus, as an independent check, in the following we derive the expression for stochastic EP using the definition in terms of probabilities of time-forward and 
time reversed trajectories.

\subsection{From path probabilities}
Now, we independently obtain the expression for stochastic EP using probabilities of time forward and time reversed trajectories.
Consider the time evolution of an ABP from $t=0$ to $\t_0$ through a path defined by $X=\{x(t), v(t), f(t) \}$. 
The motion on this trajectory involves coupling of the particle dynamics with a Langevin heat bath, and the presence of 
a non-linear self propulsion force $F(v)$. Microscopic reversibility means the probability of such a trajectory is the same
as the probability of the corresponding time-reversed trajectory. Entropy production requires break down of such microscopic reversibility.

Let us first consider the transition probability 
$p_i^+(x',v',t+\d t | x,v,t)$ for an infinitesimal section of the trajectory evolved 
during a time interval $\d t$, assuming that the whole trajectory is made up of $i=1,\dots,N$ segments such that $N \d t = {\t_0}$. 
The Gaussian random noise at $i$-th instant is described by $P(\eta_i) = (\d t/4\pi D_0)^{1/2} \exp(-\d t \, \eta_i^2/4 D_0)$.
The transition probability is given by 
\bea
p_i^+ &=&  J^+_{\eta_i, v_i} \la \d(\dot x_i - v_i) \d(\dot v_i - {\cal F}_i ) \ra \nn\\ 
&=& J^+_{\eta_i, v_i} \int d \eta_i P(\eta_i)  \d(\dot x_i - v_i) \d(\dot v_i - {\cal F}_i ),
\eea 
where the total force acting on the particle at $i$-th instant of time is ${\cal F}_i =  \eta_i + [g(v_i) +\xi(v_i)] -\p_{x_i} U(x_i) + f_i$, with
$g(v_i) = F(v_i) - \g v_i$, and the Jacobian of transformation (see Appendix-\ref{trajectory}) 
\bea
J^+_{\eta_i, v_i} = \f{1}{\d t } \left [ 1-\f{\d t}{2}\, \p_{v_i} \{g(v_i) +\xi(v_i)\} \right ].
\eea 
Thus we have $p_i^+ = J^+_{\eta_i, v_i} (\d t/4\pi D_0)^{1/2}  \d(\dot x_i - v_i) \exp[-\f{\d t}{4 D_0} \{ \dot v_i - g(v_i) - \xi(v_i) + \p_{x_i} U(x_i) - f_i \}^2] $.
The probability of full trajectory is ${\cal P}_+  = \prod_{i=1}^N p_i^+$.

Reversing the velocities gives us the time reversed path  $X^\dagger=\{x'(t'), v'(t'), f'(t') \} =\{x({\t_0}-t),-v({\t_0}-t), f({\t_0}-t)\}$, 
the probability of which can be expressed as ${\cal P}_-  = \prod_{i=1}^N p_i^-$ where 
\begin{align}
& p_i^- = J^-_{\eta_i, v_i} (\d t/4\pi D_0)^{1/2} \d(\dot x_i - v_i) \times \nn\\
&\exp[-\f{\d t}{4 D_0} \{ \dot v_i + g(v_i) -\xi(v_i) + \p_{x_i} U(x_i) - f_i \}^2], 
\end{align}
since  
$g(-v_i) = - g(v_i)$ and $\xi(-v_i) = \xi(v_i)$. The Jacobian along reverse trajectory is
\bea
J^-_{\eta_i, v_i} = \f{1}{\d t } \left [ 1-\f{\d t}{2}\, \p_{v_i} \{g(v_i) - \xi(v_i)\} \right ].
\eea 

Linearizing for small $\d t$, the ratio of the forward and backward Jacobian $J^+_{\eta_i, v_i} /J^-_{\eta_i, v_i} \simeq [ 1 - \d t (\p_{v_i} \xi) ] \simeq \exp[ - (\p_{v_i} \xi)\, \d t ] $.
The ratio of probabilities of the forward and reverse trajectories is 
\bea
\f{{\cal P}_+}{{\cal P}_- } &=& \prod_{i=1}^N e^{-\p_{v_i} \xi\, \d t} e^{(\d t /D_0) ( \dot v_i  + \p_{x_i} U -f_i - \xi(v_i) ) g(v_i) } \nn\\
&=&e^{-\int_0^{\t_0} dt \p_v \xi}\,\, e^{ {\frac{1}{D_0} \int_0^{\t_0}  dt  \left( \dot v  + \p_x U -f(t) - \xi(v_i)\right)} g(v) }.   
\label{path_ratio}
\eea
The reservoir EP over time $\t_0$ is given by $\D s_r = \kb \ln ( {\cal P}_+/ {\cal P}_- )$. Therefore, the rate of EP $\dot s_r$ gives the same expression as in
Eq.(\ref{sr_dot}). This is the first main result of our paper. Remember that $g(v) = -\g v + \zeta(v)$ is a odd function of velocity.
Assuming the initial and final steady state distributions as $P^i_s$ and $P^f_s$ respectively, the system entropy change is $\D s = s_f - s_i = \kb \ln(P_s^i/P_s^f)$.

\subsection{Entropy and dissipated heat}
\label{conservation}

The Langevin equation describing ABPs directly leads to stochastic energy balance. 
Multiplying Eq.(\ref{lange}) by velocity $v$  one obtains~\cite{Sekimoto1998} 
\bea
\dot E =  \dot W + \dot q, 
\label{1st}
\eea 
where $\dot E$ denotes the rate of change in mechanical energy  $ E= (1/2) v^2 + U(x)$, 
$\dot W =  v. f(t)$ is the rate of work done on the ABPs by external force $f(t)$, 
and $\dot q = \dot Q + \dot Q_m$ the total power absorbed by the mechanical degrees of freedom of the ABPs:
(a) from the Langevin heat bath $ \dot Q= v. (-\g v + \eta) $, and  (b) from the self-propulsion
mechanism $\dot Q_m =  v. F(v)$ with $F(v)=\zeta(v)+\xi(v)$. 

In a system of conventional passive Brownian particles, the stochastic entropy production in any process has two components. One is the rate of entropy change in the 
system $\dot s$ where the stochastic system-entropy is expressed as $s = -\kb \ln P_s$ with $P_s$ denoting steady state distribution. The 
other contribution comes from the change in entropy in the heat-bath, $\dot s_r = -\dot Q/T$~\cite{Seifert2005}. 
However, as we show below, $\dot s_r $ for ABPs has further
extra contributions coming from the mechanism of active force generation and its coupling to the mechanical forces. 

Using the Langevin equation, the reservoir EP of Eq.(\ref{sr_dot}) may be written as
\beq
\f{1}{\kb}\dot s_r = - \p_v \xi(v) + \f{g(v)}{D_0}  [ \eta + g(v) ]. \nn
\eeq
Now, $g(v) [ \eta + g(v) ] = [-\g v + \zeta(v) ]   [ -\g v + \eta + \zeta(v) ] 
=-\g \dot Q +  \zeta(v)  [\zeta(v) - 2 \g v + \eta] $. Using Langevin equation, one may replace the second term in rhs of last expression
$\zeta(v) - 2 \g v + \eta = \dot v -\g v - [f(t) - \p_x U + \xi(v)]$.  Writing $\zeta(v) = - \p_v \psi(v)$, $\zeta(v) \dot v  = - \dot \psi(v)$. 
Note that $\zeta(v) v$ is related to $\dot Q_m$, but they are not the same in presence of even function $\xi(v)$. The last term can be expressed as
\beq
\g \dot Q_{em} = \zeta(v)  \cdot [f(t) - \p_x U + \xi(v)],
\eeq
 a product of the odd part of velocity dependent force $\zeta(v)$, and all other forces that are even under time reversal. Thus, finally one obtains
 \bea
 \dot s_r =  - \f{\dot Q}{T} -\f{\zeta(v) v}{T} - \f{ \dot \psi(v)}{\g T} - \f{\dot Q_{em}}{T}  - \kb\, \p_v \xi(v).
 \label{sr_dot2}
 \eea 
 This relation clearly shows that EP in environment has several other contributions apart from the Clausius like dependence on dissipated heat $-\dot Q/T$. 
 All the other contributions appear from the internal energy source which transduce energy to mechanical motion, and cross-coupling of this process with mechanical forces.  
 This is a purely non-equilibrium effect arising due to non-linear velocity dependent self propulsion forces. 
 It is interesting to note that, this EP
 has a dependence on the energy pumped from the odd part of non-linear velocity dependent force $- \xi(v) v /T$ but not not on the total $-\dot Q_m$, 
 a term one would have naively expected if  $\dot Q_m$ could be interpreted as energy flow to the mechanical degrees of freedom from the internal depot.

 Note that if the velocity dependent force is purely an odd function of velocity, like in the case of Rayleigh-Helmholtz model and energy depot model,
 $\xi(v)=0$. In that case $\zeta(v) v = \dot Q_m$, and one gets a simpler relation~\cite{Ganguly2013} 
 \bea
 \dot s_r =  - \f{\dot Q + \dot Q_m}{T} - \f{\dot \psi(v)}{\g T} - \f{\dot Q_{em}}{T}.  
 \eea
The excess EP is due to terms not appearing in stochastic energy balance. 
Recent studies on stochastic spin dynamics showed excess EP due to rotational motion that does not contribute to energetics~\cite{Bandopadhyay2015,
Bandopadhyay2015a}. 

It is clear from the  discussions above that the definition of stochastic heat flux is directly derivable from the Langevin equation, and need not to explicitly 
refer to the time reversal parity of the dofs. In contrast, expression of stochastic EP is inherently dependent on time reversibility of the dofs. 
This happens through identification of the dissipative part of probability currents, or the structure of 
probability distributions of time reversed trajectories that explicitly depend on time reversibility of corresponding dofs. 
Physically this is expected from any entropy measure as EP quantifies the amount of breaking of time reversal symmetry.
As is seen above, all the heat flux terms $\dot Q$, $\dot Q_{m}$  and $\dot Q_{em}$ turn out to be dissipative, as well.
 While a Clausius like relation between entropy production and heat dissipation is possible at our near equilibrium, far from equilibrium
Our detailed calculations presented above shows clearly how {\em excess} entropy, added on top of the Clausius like contribution, plays an important role in the
stochastic thermodynamics of ABPs.  
 
\subsection{Fluctuation theorem}
 Eq.(\ref{path_ratio}) can be written as
$\f{{\cal P}_+}{{\cal P}_- } = \exp(\D s_r/\kb)$, 
where  $\D s_r = \int_0^{\t_0} dt \, \dot s_r$ with $\dot s_r$ given by  Eq.(\ref{sr_dot2}).  
The probability distribution of the forward process is ${\cal P}_f = P_s^i {\cal P}_+$, and that of the reverse process
is ${\cal P}_r = P_s^f {\cal P}_-$. Thus 
\bea
{\cal P}_r / {\cal P}_f = \exp(-\D s_t/\kb),
\label{eq:ratio}
\eea
with $\D s_t = \D s + \D s_r$.
This leads to the integral fluctuation theorem~\cite{Kurchan2007} 
$\la \exp(-\D s_t/\kb) \ra = \int {\cal D}[X] {\cal P}_f  \exp(-\D s_t/\kb) = \int {\cal D}[X] {\cal P}_f \, ({\cal P}_r / {\cal P}_f) =1$, which 
 readily implies a positive  entropy production on an average $\la \D s_t \ra  \geq 0$, consistent with Eq.(\ref{stdot}) and the second law of thermodynamics.
Eq.(\ref{eq:ratio}) leads to the detailed fluctuation theorem for the probability distribution of entropy production $\r(\D s_t)$~\cite{Crooks1999},
\bea
 \f{\r(\D s_t)}{\r(- \D s_t)} = e^{\D s_t/\kb},
\eea 
where $\D s_t$ denotes an amount of total entropy produced over a time interval $\t_0$. In deriving the above result it is assumed that the final distribution of the forward 
process is the same as the initial distribution of the reverse process, and vice versa -- an assumption valid in steady state.  

\subsection{Detailed balance} 
Note that at equilibrium $\dot s_t =0$ requiring $\jdd = 0$, and then the steady state condition reduces to $\nabla.\jr = 0$.  These two conditions constitute the 
detailed balance. 
The condition $\jdd = 0$ implies 
\bea
\p_v P(x,v) = \f{g(v)}{D_0} P(x,v) 
\label{jd0}
\eea
with a solution 
\bea
P(x,v) = p(x) \exp[ - \phi(v)/D_0] 
\label{pxv1}
\eea
where
$\phi(v)$ is a velocity dependent potential such that $g(v) = -\p_v \phi(v)$. 
The other condition $\nabla.\jr = 0$ can be written as,
\bea
v \p_x P(x,v) +  \p_v [ (\bF + \xi (v) ) P(x,v)]=0
\label{jr0}
\eea
in which using Eq.(\ref{pxv1}) one obtains a solution
\bea
p(x) = p_0 \exp\left[-\f{1}{v}  \int dx\,  \left(\f{g(v)}{D_0} [ \bF + \xi(v)] + \p_v \xi \right) \right].
\label{px1}
\eea
If the even function of velocity $\xi(v)=0$, and the force $\bF$ is conservative $\bF = -\p_x U$, the solution has a normalizable form $p(x)=p_0 \exp(U(x)\,g(v)/vD_0)$.
For passive particles one gets $g(v)=-\g v$ and $\xi(v)=0$ leading to the Boltzmann distribution $p(x) = p_0 \exp(-U(x)/\kb T)$. However, for an active particle the odd function 
of velocity $\zeta(v)$ is non-linear, and in general $\xi(v)$ does not vanish. Therefore, $p(x)$ is not normalizable even when $\bF=0$, not allowing detailed balance to be satisfied. Note that this conclusion is directly related to the non-zero EP even in absence of $\bF$. 

The solution given by Eq.s (\ref{pxv1}) and (\ref{px1}) satisfies Eq.(\ref{jd0}), 
if 
\bea
\f{g(v)}{D_0 v} U(x) - \f{x}{v} \left[ \f{1}{D_0} g(v) \xi(v) + \p_v \xi \right] = h(x),
\label{hx}
\eea
where $h(x)$ is entirely a function of $x$. In presence of $U(x)$, this condition can be satisfied
only if $g(v) \sim v$ and $(1/v) \p_v \xi + (1/D_0) \xi(v) =c'$, a constant. It can be easily verified that the solution of the last differential equation 
is $\xi(v) = c' \sqrt{\pi D_0/2}$ Erfi$[v/\sqrt{2 D_0}] \exp[-v^2/2D_0]$, which obeys $\xi(0)=0$, but is {\em not} an even function of $v$ due to the imaginary 
error function Erfi, violating the basic assumption regarding $\xi(v)$. 
The detailed balance condition can still be satisfied only if $c'=0$, i.e., $\xi(v)=0$. Under this condition it is easy to see that 
Eq.(\ref{hx}) is trivially satisfied with $g(v) \sim v$, which denotes equilibrium for {\em passive} particles up to a scaled temperature, and a Maxwell-Boltzmann velocity
distribution.

\subsection{Free Rayleigh-Helmholtz particle: Apparent detailed balance and internal EP}
It was shown in Ref.~\cite{Chaudhuri2014} that a free Rayleigh-Helmholtz (RH) particle, in absence of external force or spatial potential profile, obeys detailed balance,
although evidently is a non-equilibrium system with activity maintained by a velocity dependent force. The corresponding steady state distribution  
$P_s(v) = {\cal N} \exp [ -\phi(v)/D_0]$, where $\phi(v) = (\g/2) v^2 - (a/2) v^2 + (b/4) v^4$ with $a>\g$, is also unlike the equilibrium Maxwell-Boltzmann distribution.
The system obeys detailed balance in velocity space, and produces no entropy. However, a self-propelled RH particle being far from equilibrium, must produce entropy because of its self propulsion.
This fact could not be captured within the RH model itself, as it does not involve any explicit mechanism behind self-propulsion. In order to get a better insight, here we consider
a model with an internal energy depot, having energy $e(t)$ that evolves as~\cite{Schweitzer1998} 
\beq
\f{de}{dt} = \dot q_e - r_m e - \nu(v) e.
\label{en}
\eeq 
Here $q_e$ is a rate of energy gain by the energy depot, via nutrient intake by a living organism, $r_m$ is the metabolic rate required to maintain the organism alive, 
and $\nu(v) e$ is  a rate of energy dissipation towards its motility. The Langevin equation of motion in absence of external force or potential is
\bea
\dot x &= & v, \nn\\
\dot v &=& -\g v + \eta(t) + \zeta(v).  
\label{lange_rh}
\eea
Note that $\nu(v) e = v \zeta(v)$ is the energy dissipated from the internal depot to the motion of the ABP. 
The Fokker-Planck equation for the joint probability distribution $P(e,x,v,t)$ is given by
\bea
\p_t P(e,x,v,t) &=& - \p_e [(\dot q_e - r_m e - \nu(v) e) P ] - \p_x (vP) \nn\\
          && - \p_v [ g(v) P ]  + D_0 \p_v^2 P \equiv \nabla.\jv,
\eea
where $g(v) = -\g v + \zeta(v)$. Under time reversal $\dot q_e$ and $r_m$ are assumed to be odd, 
and $e$ is an even parity variable. Since $\zeta(v)$ is odd, $\nu(v)$ is also an even 
parity variable. The last step above denotes $\nabla \equiv (\p_e, \p_x, \p_v)$, and $\jv \equiv (j_e, j_x, j_v)$. The probability current may be decomposed into 
a time reversible $\jv_r \equiv [ (\dot q_e - r_m e) P,\, v P,\, 0 ] $, and a dissipative part $\jv_d \equiv [j_d^e, j_d^x, j_d^v] \equiv [ -\nu(v)e\,P,\, 0,\, g(v)P - D_0 \p_v P ]$. 

Thus detailed balance condition $\jv_d=0$, including the internal activity producing mechanism, 
requires $\nu(v) e = \zeta(v) v =0$, i.e., self propulsion force $\zeta(v)=0$. This along with
$g(v)P - D_0 \p_v P = 0$ leads to $D_0 \p_v P = - \g v P$. This has the equilibrium solution $P(v) = {\cal N} \exp(-v^2/2 \kb T)$. 

Using the same method as in Sec~\ref{ep_fp1}, one can then proceed to obtain the stochastic EP in the reservoirs. Denoting the extended phase space integral 
by $d\o = de\, dx\, dv$, the average EP of the system is given by $\dot S = \kb \int d\o \ln P [\nabla. (\jv_r + \jv_d)]$. After a little algebra one obtains 
$\int d\o  \ln P \nabla.\jv_r= - \la r_m \ra$ where $\la r_m \ra = \int d\o P r_m$. On the other hand, 
$\int d\o \ln P \nabla.\jv_d  = - \la \nu(v) \ra + \int d\o [j_d^v - g(v) P] [j_d^v/P D_0]$. Thus the stochastic EP in the reservoir can be expressed as
\bea
\f{\dot s_r}{\kb} = r_m + \nu(v) + \f{\dot v g(v)}{D_0}.
\eea
The last term on the right hand side is same as the terms derived in Eq.(\ref{sr_dot}) for free ABPs in absence of the even function of velocity. 
The other two terms occur due to explicit consideration of the energy depot mechanism to produce self propulsion. Note that, even if 
the particle does not produce active velocity dependent forces, with energy dissipation to motion $\nu(v)=0$, this model predicts stochastic EP 
in terms of the metabolic rate  $r_m$, that keeps the organism alive. However, it is interesting to note that the terms due to the fast internal dofs,
$r_m$ and $\nu(v)$ do not get coupled non-trivially with the slow modes, unlike the emergence of cross terms between slow modes like odd and even functions
of velocity giving rise to $\dot Q_{em}$. 

Assumption of a faster time scale for getting the steady state of the internal energy depot, 
$de/dt=0$, gives $e = \dot q_e/[r_m + \nu(v)]$ leading to $\zeta(v) = \dot q_e \nu(v)/[v r_m + v \nu(v)]$. Assuming
$\nu(v) = c v^2$, one gets $\zeta(v) =  v \dot q_e/[r_m + c v^2] \approx a v - b v^3$ corresponding to the RH model, 
with $a=\dot q_e/r_m$, $b=\dot q_e c/r_m$ in the limit of $v^2 \ll r_m/c$.
Further detailed study of ABP models including internal mechanism for self propulsion will be presented elsewhere~\cite{Chaudhuri_prep}.


\begin{figure}[t]
\begin{center}
\includegraphics[width=7 cm] {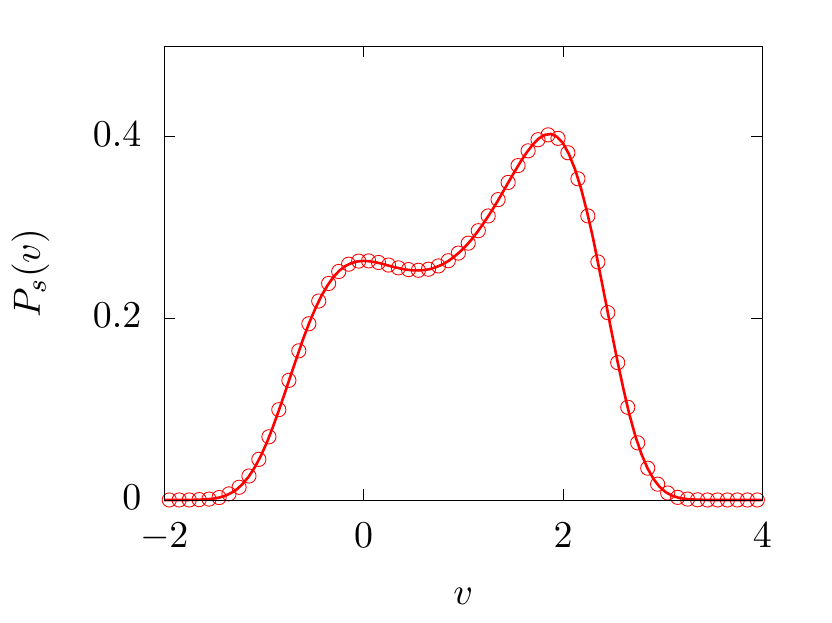}
\caption{(Color online) 
Steady state probability distribution obtained from simulation (points), compared  against the  line  drawn using the analytic form
$P_s(v) = {\cal N} \exp[-\chi(v)/D_0]$ with $\chi(v) =  \f{1}{2}(a+\g) v^2 - \f{b}{3} v^3 + \f{c}{4} v^4$ where $a=0$,
$\g=1$, $b=2.4$, $c=1$ and $D_0=1$. 
}
\label{fig:pv}
\end{center}
\end{figure}
%


\subsection{Probability distribution of entropy production}
Let us now return to the coarse grained ABP model containing only velocity dependent forces, and 
consider a velocity dependent potential $\chi(v) =  \f{1}{2}(-a+\g) v^2 - \f{b}{3} v^3 + \f{c}{4} v^4$ such that the velocity dependent force $F(v) = g(v)+\xi(v) = -\p_v \chi$. 
The corresponding Langevin equation of  motion under this force  
$\dot v = \eta + g(v) + \xi(v)$ with $g(v) = (a-\g) v  - c v^3$ and $\xi(v)= b v^2$. 
At steady state, the mean velocity has three solutions, $v=0, (b/2c) \pm \sqrt{b^2-4(-a+\g)c}/2c$. Among these solutions $v=0$ and $v=(b/2c) + \sqrt{b^2-4(-a+\g)c}/2c$ are stable fixed points and $v=(b/2c) - \sqrt{b^2-4(-a+\g)c}/2c$  is an unstable fixed point. The non-zero velocity stable fixed point
gets viable for $b^2 \geq 4c(-a+\g)$. 
In absence of external potential or force, the probability distribution is independent of position, obeying the FP equation  
$\p_t P(v) = \p_v [ P \p_v \chi + D_0 \p_v P]$. This has a steady state solution $P_s = {\cal N} \exp[-\chi(v)/D_0]$ carrying non-zero dissipative current $\jdd = g(v)P-D_0\p_v P$.

From Eq.(\ref{sr_dot2}), the EP in the reservoir over time $\t_0$ is expressed as
$\D s_r = - \f{ 1}{T}\left [\D Q + \f{\D \psi }{\g} + \D Q_{em} \right]  
-\int^{\t_0} dt \,  \left[ \f{1}{T} \zeta(v) v  + \kb \p_v \xi(v) \right],$  
where $\D Q $ is the heat absorbed over $\t_0$, $\psi(v) = -\int dv\, \zeta(v) = a v^2/2 + c v^4/4$, $\D Q_{em} = \int^{\t_0}  dt\, \zeta(v)\, \xi(v)$. 
The simplest possible choice of such active velocity dependent force is $F(v) = b v^2 - c v^3$, with  $a=0$ and $b \geq \sqrt{4 c \g}$ such that a real $v=(b/2c)+(b^2-4\g c)^{1/2}/2c$ stable fixed point in velocity is available.
Note that the system EP over  time $\t_0$ is  $\D s = \kb \ln [P_s(\t)/P_s(0)] = \D \chi(v)/\g T$ where $\D \chi = \chi(v (\t)) - \chi(v(0))$, and 
$\chi(v) = \hf \g v^2 - \f{b}{3} v^3 + \f{c}{4}v^4$.
Moreover, energy conservation, as discussed in Sec.\ref{conservation}, implies $\D Q = \D E - \D Q_m$ 
as the work done due to external force is zero  in the case considered here.  
Therefore, the total EP over time $\t_0$ is $\D s_t = \D s + \D s_r = -(1/T) [\D E -\D Q_m + \D Q_{em}] - \int^{\t_0} dt [\zeta(v) v/T  + \kb \p_v \xi(v)] - (1/\g T) (\D \psi - \D \chi)$.
Using $a=0$, one obtains $\D E + (\D \psi - \D \chi)/\g = \D (b v^3/3\g)$, as $E=v^2/2$. 
Note that $-\D Q_m + \D Q_{em} + \int dt \zeta(v) v dt =\int dt [-v F(v) + \zeta(v) \xi(v) + \zeta(v) v] = \int dt\, [-v + \zeta(v)]\, \xi(v)$.
Thus, one obtains 
\bea 
\D s_t 
&=& -  \f{1}{T} \left [ \D \left(\f{b }{3 \g} v^3 \right)+ \int^{\t_0} dt \{ \zeta(v) - v\} \xi(v) \right] \nn\\
&& - \int^{\t_0} dt\,  \kb \p_v \xi(v).
\label{Dst}
\eea

%
 \begin{figure}[t]
\begin{center}
\includegraphics[width=7 cm] {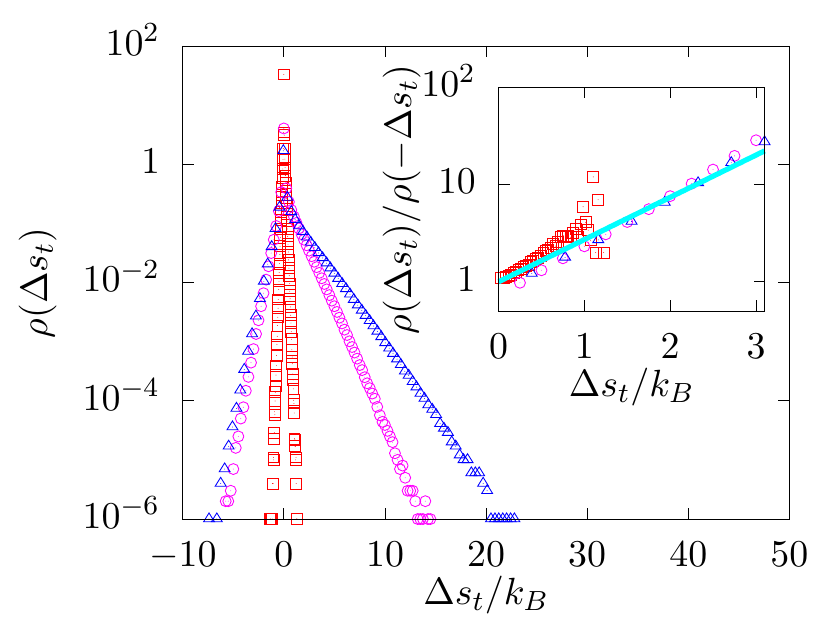}
\caption{(Color online) 
Probability distribution of entropy production $\D s_t$ over time span $\t_0$ = 16 ($\Box$), 64 ($\circ$), 128 ($\triangle$)\, $\d t$ 
plotted in linear-log scale. 
Inset: Ratio of probability distribution of positive and negative entropy production in linear-log scale. The solid line shows the function $\exp(\D s_t/\kb)$. The deviation of data from this line
is due to lack of statistics at large $\D s_t$.
}
\label{fig:ps}
\end{center}
\end{figure}


We  perform numerical integration of the Langevin dynamics of this ABP using Stratonovich discretization with time step $\d t = 10^{-4} \t$, where $\t=1/\g$, 
and parameters $D_0=1$, $\kb T=1$, $b=2.4$ and $c=1$. 
Figure~\ref{fig:pv} shows a plot of steady state velocity distribution obtained from the numerical simulation, showing good agreement with the analytic expression $P_s = {\cal N} \exp [ -\chi(v)/D_0]$.
The distribution function has two maxima, at $v=0$ and $v=(b/2c) + \sqrt{b^2-4\g c}/2c$. On an average, the ABP moves towards the positive $x$ axis.
From this simulation, we further obtain probability distribution of total stochastic EP, $\D s_t$, using the expression in Eq.~\ref{Dst}, over different time spans $\t_0$. 
The distribution function has a sharp peak at $\D s_t=0$, but gets broader for longer observation time $\t_0$ (Fig.\ref{fig:ps}). As shown in the inset of Fig.~\ref{fig:ps}, the ratio of probability distribution of positive and negative 
EP shows good agreement with the detailed fluctuation theorem $\ln [\r(\D s_t)/\r(-\D s_t)] = \D s_t/\kb$.

\section{Discussion}   
Models of self propelled particles in presence of non-linear velocity dependent force have been studied extensively in recent literature. 
We have shown earlier, if the velocity dependent force is odd under time reversal, the ABPs can not produce entropy  
unless coupled to conservative or non-conservative  force~\cite{Ganguly2013, Chaudhuri2014}. Given the self propulsion of the particles, 
even free ABPs should have produced entropy. 
In this paper, using an internal mechanism for generation of self propulsion, namely, in terms of an internal energy depot, we have shown, free ABPs of RH kind indeed
produce entropy, albeit via the internal mechanism of producing self-propulsion, keeping the expression for EP in the velocity space unaltered. After integrating out the 
faster internal dofs, the self propulsion turns up as non-linear velocity dependent force in ABPs spatial motility.

We studied such coarse-grained models of ABPs, without explicit mechanism of self-propulsion, 
in the presence of a generic non-linear velocity dependent force, containing both odd and 
even functions of velocity. This leads to autonomous entropy production in velocity space. 
We have derived the expression for the total EP, independently, using the Fokker-Planck equation, and probability of time forward and time-reversed trajectories. Both the methods led to the same result. Note that the Jacobians in the path probability method, corresponding to the time forward and time reversed trajectories are not the same, unlike other simpler systems~\cite{Narayan2004,Seifert2012,Seifert2008}. In fact, the ratio of these Jacobians contributes to the dependence of EP on the even part of the velocity dependent force under time reversal. 

The total stochastic EP obeys fluctuation theorems. It is interesting to note that the EP in the reservoir has several excess contributions in addition to the Clausius entropy related to dissipated heat. This excess entropy shows two fundamentally new contributions with respect to earlier study involving force that is only an odd function of velocity~\cite{Chaudhuri2014}. These are: (i)~a velocity gradient of the even function of velocity, and (ii)~cross-coupling of the odd and even functions of velocity. 
Using numerical simulations we have obtained probability distribution of the total EP and found good agreement with detailed fluctuation theorem. 
Note that the observation of { excess} EP,  does not have any conflict with thermodynamic inequality due to Clausius, for systems out of equilibrium. 
In conclusion, each non-equilibrium dissipative mechanism not only adds to entropy independently, they often couple with each other to give rise to new terms in EP.

\acknowledgements
We thank A. M. Jayannavar and Abhishek Dhar for discussions, Swarnali Bandopadhyay for a critical reading of the manuscript and useful comments. Financial support from SERB, India  is gratefully acknowledged.

\appendix
\section{Probability of a trajectory}
\label{trajectory}
The Langevin dynamics is described by
 \bea
 \dot x &=& v \nn\\
 \dot v &=& F(v) + \eta(t) + \F 
 \eea
 where $F(v) = g(v)+\xi(v)$ with $g(v)=-\g v + \zeta(v)$, and $\F$ denotes the velocity-independent forces.
 The Gaussian white noise is characterized by
$\la \eta(t) \ra =0$, $\la \eta(t)\eta(0) \ra = 2 D_0 \d(t)$ with $D_0= \g \kb T$.
 Discretizing the equation with $t=i\, \d t$, using Stratonovich rule, 
\bea
x_{i} &=& x_{i-1} + \f{1}{2} (v_i + v_{i-1}  )\d t \nn\\
v_i &=& v_{i-1} + \f{1}{2} \left[  F(v_i) + \F(x_i) +  F(v_{i-1}) + \F(x_{i-1})\right] \nn\\
&& + \eta_i \d t.
\label{apt}
\eea
The Gaussian random noise $\eta(t)$ follows the distribution $P(\eta_i) = (\d t/4 \pi D_0) \exp(-\d t \eta_i^2/4 D_0)$
where $D_0 = \g \kb T$. The transition probability over $i$-th segment of the trajectory  
$p^+_i \equiv P(x_i, v_i | x_{i-1},v_{i-1}) = J_{\eta_i,v_i} \,\la \d (\dot x -v) \d (\dot v - \{F(v) +\F \} ) \ra $
leads to
 \bea
 p^+_i = J_{\eta_i,v_i}  \,\d (\dot x -v) \sqrt{\f{\d t}{4 \pi D_0}} e^{-\f{\d t}{4 D_0} \left[ \dot v -F(v) -  \F \right]^2 },
 \eea
 where
 \bea
 J_{\eta_i,v_i} &=& {\rm det} \left( \f{ \p \eta_i}{\p v_i }\right) =\f{1}{\d t} \left( 1 -\f{\d t}{2} \p_{v_i} F(v_i)\right)\nn\\
 &=& \f{1}{\d t} \left( 1 -\f{\d t}{2} \p_{v_i} [ g(v_i) + \xi(v_i) ]\right)
 \eea
 using Eq.(\ref{apt}).
The probability associated with a full trajectory is ${\cal P}^+ = \prod_i p^+_i$.

\bibliographystyle{prsty}

\end{document}